\documentclass[10pt, twocolumn]{article}

\usepackage{geometry}
\usepackage{amsmath}
\usepackage{graphicx}

\newcommand\Atitle[1]{\Large \bf \noindent \begin{center} #1
\end{center}\rm \normalsize \vskip.125in }%

\newcommand\Aauthor[1]{\vskip.1in\noindent%
   \large  \begin{center} \textsf{#1} \end{center}\rm \vskip-.2in}

\newcommand\ocis[1]{\vskip-1pc \hskip3pc {\normalsize\it OCIS codes: \rm #1 \hfill}\vskip1pc}

\renewenvironment{abstract}
{\vskip2.75pc\noindent\begin{center}\begin{minipage}{5.5in} \parindent.2in
   \noindent \rm}{\hskip.07in \copyright \hskip.05in 2009 \hskip.05in
   Optical Society of America \\ \hfil \end{minipage}\end{center}}

\newenvironment{abstract*}
{\vskip2.75pc\noindent\hskip.3in\begin{minipage}{5.5in} \parindent.2in
   \noindent \rm}{\hfil \end{minipage}\vskip.25in}

\let\title\Atitle
\let\author\Aauthor

\let\address\Aaddress

\renewcommand\maketitle{} 

\begin{document}

\twocolumn[  

\title{Quantitative phase reconstruction for orthogonal-scanning\\ 
differential phase-contrast optical coherence tomography}

\author{Bettina Heise$^{1,2,3,*}$, David Stifter$^{3}$}

\address{
$^1$ RECENDT GmbH, Hafenstrasse 47-51, 4020 Linz, Austria\\
$^2$ FLLL - Department of Knowledge-based Mathematical Systems,\\ Johannes Kepler University, Altenberger Strasse 69, 4040 Linz, Austria\\
$^3$ ZONA - Center for Surface and Nanoanalytics, \\Johannes Kepler University, Altenberger Strasse 69, 4040 Linz, Austria\\
$^*$Corresponding author: Bettina.Heise@jku.at
}
 
\begin{abstract}
\noindent We present differential phase-contrast optical coherence tomography (DPC-OCT) with two transversally separated probing beams to sense phase gradients in various directions
by employing a rotatable Wollaston prism. In combination with a two-dimensional mathematical reconstruction algorithm based on a regularized shape from shading (SfS) method accurate quantitative phase maps can be determined from a set of two orthogonal en-face DPC-OCT images, as exemplified on various technical samples.  
Original paper: Optics Letters, Vol. 34, Issue 9, pp. 1306-1308 (2009);
http://dx.doi.org/10.1364/OL.34.001306
\end{abstract}

\ocis{110.4500, 120.3180, 120.5050, 110.3010, 120.4290}
]  

\noindent Optical Coherence To\-mo\-gra\-phy  (OCT), a depth-resolved imaging method for the characterisation of turbid media, has become a well-established technique in biomedical diagnostics and research. In case of transparent and weakly scattering specimens - like single cell layers - it was shown that enhanced contrast can be obtained by combining the OCT principle with that of differential interference contrast (DIC) microscopy, resulting in differential phase contrast OCT (DPC-OCT)~\cite{hitzenberger_ol99}. With DPC-OCT, sub-wavelength differences in optical path lengths (OPL) could be quantitatively measured~\cite{sticker_ol01,rylander_ol04}, even below extended scattering layers. However, the phase difference is determined between two adjacent probing beams, leading to unidirectional image contrast along the direction of beam separation and leaving features with phase gradients perpendicular to the separation direction undetected.
In addition, the unidirectional contrast in the images unavoidably causes severe streaking artifacts when quantitative phase maps are obtained by numerical line integration, as proposed in~\cite{rylander_ol04}, due to noise accumulation and undetermined integration constants. These artefacts can be avoided if a well defined flat reference plane is available close to the structures of interest: a collinear, phase-sensitive approach in the time-domain (TD) was used to determine the phase difference between such an additional reference plane and the sample itself~\cite{rylander_ol04}. However, this technique requires that even for the evaluation of a single sample layer a whole 3D volume - including both, layer and reference plane - has to be acquired. 

For the recently developed phase-sensitive spectral-domain (SD) OCT techniques a flat reference plane is equally placed in the vicinity of the sample structures to form a stable common path interferometer~\cite{choma_ol05,joo_ol05}. Although these SD-methods outperform their TD-counterparts~\cite{hitzenberger_ol99,sticker_ol01,rylander_ol04} in terms of phase stability, TD-OCT techniques are still favourable in certain cases, like transversal scanning and full-field (FF) OCT~\cite{hitzenberger_oexp03,dubois_applop04} in combination with dynamic focusing leading to a constant high lateral resolution throughout the whole imaging depth. In addition, phase retrieval with SD-techniques requires the acquisition, calculation and evaluation of whole 3D-data sets: three orders of magnitude more data is needed for SD-OCT compared to TD-OCT in order to obtain an en-face scan (C-scan) of a single interface. Consequently, acquiring individual C-scans is still 5-10 times faster with TD-techniques~\cite{hitzenberger_oexp03} despite the high acquisition rates of current SD methods~\cite{potsaid_oexp08}.

We present a novel method which combines orthogonal scanning DPC-OCT in the TD with a two-dimensional phase reconstruction approach and which is fully compatible with high-speed transversal scanning OCT as well as with FF-OCT. Our method effectively eliminates streaking artefacts so that accurate quantitative phase maps can be obtained, as exemplified on diverse technical samples, underlining also the need of enhanced contrast and quantitative information retrieval for alternative OCT applications outside the biomedical field~\cite{stifter_applphysb07}.  

\begin{figure}[htbp]
\centerline{\includegraphics[width=8.3cm]{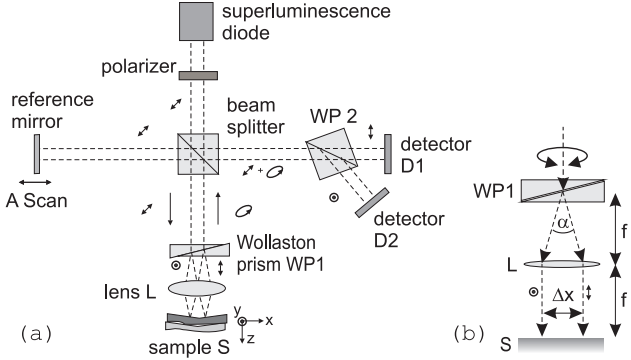}}	 
\caption{ (a) Scheme of the DPC-OCT setup with indicated polarization directions, (b) configuration of the probing head with rotatable Wollaston prism ($f$ focal length, $\alpha$ divergence angle, $\Delta x$ probing beam shear)}
\label{fig:Fig1}
\end{figure}

The DPC-OCT setup is schematically depicted in Figure~\ref{fig:Fig1} and has been realised by modifying an existing polarisation sensitive (PS)-OCT system for strain imaging in polymers and composite materials (center wavelength $\lambda$=1550~nm, 90~dB system sensitivity, 19~$\mu$m axial resolution in air)~\cite{stifter_applphysb03}: a Wollaston prism (WP1) is inserted in the sample arm in the focal point of the achromatic lens L with focal length $f=$~4.5~mm. WP1 has a divergence angle $\alpha=$~5', thus 
a lateral shear $\tau=6.5~\mu$m, determining the lateral resolution, is introduced between the two polarisation-encoded probing beams (Figure~\ref{fig:Fig1}\,(b)), as verified by calibration measurements  on a sharp reflecting edge. 
In addition, WP1 can be freely rotated around the beam axis, so that the shear can be oriented both in $x$ or $y$ direction as $\tau_x =\Delta x$ and $\tau_y =\Delta y$, respectively.
Due to refractive index changes or thickness variations an OPL difference causes a  phase shift $\Delta \varphi_i $ (with $i=x,~y$) between the two sheared beams. In the backreflected beam this generally results in an elliptical polarization state which is brought to interference with the linearly polarised light of the reference beam (oriented at $45^\circ$). The two depth ($z$) resolved interference signals $S_1(x,y,z)$ and $S_2(x,y,z)$ are detected at each lateral position $(x,y)$ in the PS-detection unit (WP2 and two photodetectors $D_1$ and $D_2$). Scaling the directional phase shift $\Delta \varphi_i$ with the beam shear $\tau_i$ yields the phase gradient $\varphi_i$,
\begin{eqnarray}
\varphi_i(x,y)= \frac{\Delta \varphi_i}{\tau_i}= \frac{\varphi_{2i}(x,y)-\varphi_{1i}(x,y)}{\tau_i}, \; i=(x,y).  
\end{eqnarray}
The two phase functions $  \varphi_{1i} $ and $  \varphi_{2i} $ of the detector signals $S_1(x,y,z)$ and $S_2(x,y,z)$ are determined by applying the Hilbert transform $\cal{H}$ \cite{sticker_ol01}  with
\begin{eqnarray} 
\varphi_{1,2 x}(x,y)=\arctan\frac{{\cal{H}} (S_{1,2}(x \pm \tau_x/2 ,y,z))}{S_{1,2}(x\pm \tau_x/2 ,y,z)} ,
\end{eqnarray}
for the beam shear in $x$-direction and for $\varphi_{1,2 y}$ introducing the shear in $y$-direction by an analogous expression.

For the estimation of the achievable accuracy of the setup, we determined the phase distribution by repeating 2500 A-scans on a simple glass slide and by evaluating the 40~dB reflection of the bottom surface. Despite the rather long interferometer arms ($>$0.5~m), the standard deviation of the phase fluctuations was determined to be only 0.0055~rad corresponding to 0.7~nm OPL at a center wavelength of 1550~nm.  
\begin{figure}[htbp]
\centerline{\includegraphics[width=8.3cm]{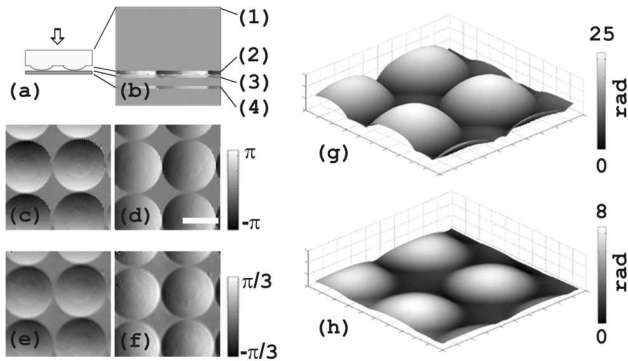}} 
\caption{(a) Sketch of a microlens array test structure with 4 interfaces, (b) cross-sectional DPC-OCT image, (c, d) en-face DPC-OCT images in $x$- and $y$-direction at interface (2), (e, f) en-face DPC-OCT images at interface (4), (g) reconstructed cumulative phase map at interface (2), (h) cumulative phase map at interface (4). White scale bar: 100~$\mu$m}
\label{fig:Fig2}
\end{figure}

In Figure~\ref{fig:Fig2} $ \Delta \varphi_i(x,y)$ maps are depicted for a microlens array (Suss MicroOptics, Switzerland) with known dimensions in order to validate our method (Figure~\ref{fig:Fig2}\,(c)-(f)). The array was placed upside down on a microscope slide to simulate several buried layers (cf. cross-sectional sketch and corresponding DPC-OCT scan in Figure~\ref{fig:Fig2}\,(a),\,(b)).
Using both orthogonal DPC images $\Delta \varphi_x(x,y)$ and $\Delta \varphi_y(x,y)$ of the individual layers, the cumulative phase map $\hat\varphi(x,y)$ can be mathematically reconstructed with a regularized shape from shading (SfS) approach \cite{17Vogel}, which we adapted  for the first time for the phase reconstruction from DPC-OCT images (Figure~\ref{fig:Fig2}\,(g),\,(h)). In this approach  the energy functional $\cal{E}$ according to  equation (3) has to be minimized with respect to the total phase $\hat \varphi$ including a penalty term with the regularization parameter $\mu $ for higher order derivatives, 
\begin{eqnarray}
{\cal{E}}( \hat \varphi(x,y))&=&\int_\Omega\left( (\textbf{f}(x,y)-\textbf{R}( \hat \varphi_x,  \hat \varphi_y)\right)^2
\nonumber
\\ 
& &  + \mu ( \hat \varphi_{xx}^2+2  \hat \varphi_{xy}^2+  \hat \varphi_{yy}^2) ,
\end{eqnarray}
with the measured data
$
\textbf{f}=
\left(
\begin{array}{c}
  \varphi_x
\\
  \varphi_y	
\end{array}
\right)
$
and  
\(
\textbf{R}=
\left(
\begin{array} {c}
 \hat \varphi_x\\
 \hat \varphi_y
\end{array}
\right)
\)
denoting the derivatives of the total phase $ \hat \varphi$ to be reconstructed. In case of the microlens array the mathematical reconstruction of the OPL map at interface (2) yields - for a sample refractive index $n_s$=1.449 at $\lambda$=1550~nm - a total geometrical height of the individual lenslets of 1.64 $\mu$m, closely matching the technically specified value of 1.63~$\mu$m. 

\begin{figure}[htbp]
\centerline{\includegraphics[width=8.3cm]{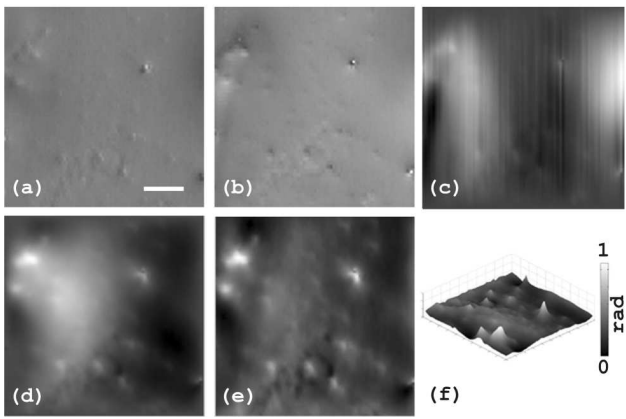}} 
\caption{DPC-OCT image of an oil layer with embedded particles. (a, b) DPC images with shear in $x$- and $y$-direction, respectively, (c) OPL reconstruction by numerical line integration using the single scan displayed in (b), (d) FFT-based OPL reconstruction from scans (a) and (b), (e, f) OPL reconstruction with SfS method using scans (a) and (b).  White scale bar: 100~$\mu$m} 
\label{fig:Fig3}
\end{figure}

In Figure~\ref{fig:Fig3} the performance of the SfS algorithm is compared to other reconstruction methods, using $  \Delta \varphi_x(x,y)$ and $\Delta \varphi_y(x,y)$ maps of the top reflection of a thin corrosion inhibiting oil layer with small embedded abrasion particles. At first, we tested straight forward numerical line integration on a single en-face DPC image (Figure~\ref{fig:Fig3}\,(b)), which leads to the above described streaking artifacts (Figure~\ref{fig:Fig3}\,(c)), especially severe for samples with low signal-to-noise ratio (SNR).
Using both DPC scans (Figure~\ref{fig:Fig3}\,(a),\,(b)) we then reconstructed the OPL map with a global Fast Fourier Transform (FFT)-based approach by solving the Laplace problem in the Fourier domain \cite{16Frankot-Chellappa}. This method performs better than line integration, especially for samples with low SNR (Figure~\ref{fig:Fig3}\,(d)). 
Nevertheless, slight blurring and haze is present, partly caused by the low pass characteristics of the inverse Laplace filtering. Boundary artefacts, inherent to FFT-based methods, can be reduced by image continuation and windowing, but will at the same time deteriorate the results for extended sample features.In comparison, reconstruction by SfS yields a significantly improved result as depicted in Figure~\ref{fig:Fig3}\,(e),\,(f) with no streaking artefacts or blurred features and areas.
 
\begin{figure}[h]
\centerline{\includegraphics[width=8.3cm] {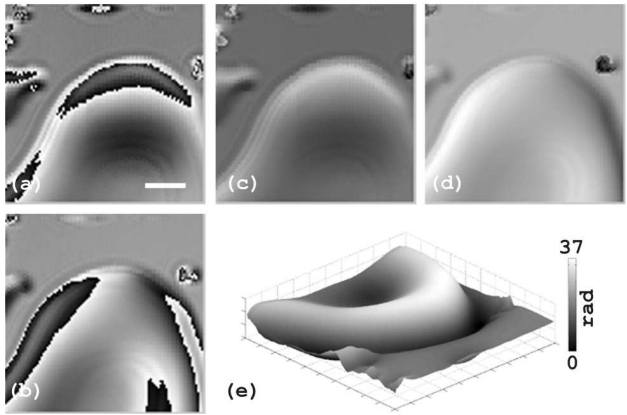}} 
\caption{(a, b) Wrapped DPC images of a transparent polymer layer (shear in $x$- and $y$-direction, respectively), (c, d) unwrapped phase of (a) and (b), respectively, (e) total phase map reconstructed from (c) and (d). White scale bar: 40~$\mu$m }
\label{fig:Fig4}
\end{figure}

As a last example we recall that the measured phase differences are mapped into the interval $[-\pi,\pi]$ and that phase unwrapping is required in case of steep phase gradients as depicted in Figure~\ref{fig:Fig4}\,(a),\,(b) for a soft and uneven acrylic polymer film.  We applied a quality guided unwrapping approach \cite{12Ghiglia} to obtain the two orthogonal unwrapped DPC en-face scans (Figure~\ref{fig:Fig4}\,(c),\,(d)). The SfS reconstruction of the unwrapped images leads to the cumulative phase distribution (Figure~\ref{fig:Fig4}\,(e)), demonstrating that this method - also in combination with phase unwrapping - acts as robust method coping with steep and extended features as well as handling low SNR images like those in Figure~\ref{fig:Fig3}.  

In conclusion, we have demonstrated TD-DPC-OCT imaging of en-face phase gradients in variable directions combined with a regularized SfS reconstruction approach to determine quantitative OPL maps: from two orthogonal shear directions a more robust solution is obtained compared to conventional reconstruction methods, like numerical integration of single DPC-OCT images. Whereas the current setup has been realised with a rotatable Wollaston prism, polarisation-dependent electro-optic beam steering could be used in the future. Furthermore, our method is fully compatible with high-speed transversal scanning as well as with FF-OCT and provides enhanced contrast and quantitative phase information for applications ranging from biomedicine to industrial metrology.
 
This work was financially supported by the Austrian Science Fund (FWF), project P19751-N20, by the European Regional Development Fund (EFRE), and the federal state Upper Austria. We thank C.K. Hitzenberger and his group (Medical University of Vienna) for continuous scientific support.

\end{document}